*Andrey Subochev, Igor Zakhlebin*

**ALTERNATIVE VERSIONS OF THE GLOBAL COMPETITIVE INDUSTRIAL PERFORMANCE RANKING CONSTRUCTED BY METHODS FROM SOCIAL CHOICE THEORY**

**Abstract**

The Competitive Industrial Performance index (developed by experts of the UNIDO) is designed as a measure of national competitiveness. Index is an aggregate of eight observable variables, representing different dimensions of competitive industrial performance. Instead of using a cardinal aggregation function, what CIP's authors do, it is proposed to apply ordinal ranking methods borrowed from social choice: either direct ranking methods based on majority rule (e.g. the Copeland rule, the Markovian method) or a multistage procedure of selection and exclusion of the best alternatives, as determined by a majority rule-based social choice solution concept (tournament solution), such as the uncovered set and the minimal externally stable set. The same method of binary comparisons based on majority rule is used to analyse rank correlations. It is demonstrated that the ranking is robust but some of the new aggregate rankings represent the set of criteria better than the original ranking based on the CIP.

This study comprises research fi ndings from the "Constructing Rankings by Social Choice methods" project (grant № 12-05-0036, years 2012–2013) carried out as a part of The National Research University Higher School of Economics' Academic Fund Program. The work was partially financed by the International Laboratory of Decision Choice and Analysis (DeCAn Lab) as a part of projects 32.0 (2010), 53.0 (2011) and 93.0 (2013) within the Program for Fundamental Research of the National Research University Higher School of Economics. We are grateful to professor F. Aleskerov for his helpful suggestions and careful review of the manuscript.

*Andrey Subochev*, DeCAn Lab and Department of Mathematics for Economics, National Research University Higher School of Economics, Moscow, asubochev@hse.ru
*Igor Zakhlebin*, National Research University Higher School of Economics, Moscow, zahl.igor@gmail.com

# 1 Introduction

National competitiveness is broadly defined as an ability of a national economy to produce goods and services that meet the requirements set by international competition, while citizens enjoy a standard of living that is both improving and sustainable [Tyson, 1992]. Although no general consensus on how to determine national competitiveness has been reached, it is agreed that this is not a self-contained notion. In order to measure it one has to define a set of factors such that their values either determine the level of national competitiveness or are determined by it. Once this set of factors has been defined, the measurement of national competitiveness becomes a problem of multiple criteria aggregation.

This paper deals with Competitive Industrial Performance Index (CIP), presented in UNIDO's Competitive Industrial Performance Report 2012/2013. The CIP Index is based on eight factors grouped into three sets called dimensions. Index value is a product of six values: two arithmetic means of two pairs of factors, which form the second dimension, and values of the other four factors. In this paper we do not question either definition of competitiveness, proposed by authors of the report, nor their choice of its observable correlates. We are interested in how the aggregation is performed.

The method of aggregation adopted by the authors of the CIP is theoretically problematic. Since the aggregation formula itself and the values of weights (factors for summations and powers for multiplication) are not unique, their choice have to be justified. It is extremely difficult if not altogether impossible to justify one's choices when the resulting variable is not directly observable and measurable. We have no such justification for the problem under consideration, therefore we cannot be sure that calculation of the CIP index presented in the report is a correct aggregation procedure yielding meaningful results. A *cardinal* value of this index will not tell us anything about performance of a given country if we do not compare it with other countries' values. The differences or proportions of index values across countries or over time have no evident interpretation as well. The only use we can make of the index is to order countries with respect to their CIP values in a given year.

As a partial solution to the problem of interpretation of cardinal values as well as another way to test the robustness of the ranking based on the CIP index we propose to apply *ordinal* ranking methods. We borrowed them from social choice theory since it is possible to frame any multi-criteria decision problem as a social choice problem. Eight industrial competitiveness factors are regarded as criteria. Countries are ranked by their values of each factor first, then eight factor-based-rankings are aggregated by simple majority rule. The result of the aggregation is a binary relation. It tells us which country from a given pair is better than the other one with respect to majority of criteria. This majority relation is intransitive generally. Therefore, in order to obtain a ranking we need to apply either a direct ranking method based on majority rule (e.g. the Copeland rule) or a multistage procedure of selection and exclusion of the best countries, as determined by a majority rule-based social choice solution concept (tournament solution), such as the uncovered set and the minimal externally stable set.

The aims of the paper are the following. First, we use ordinal methods of aggregation to produce alternative versions of the CIP ranking. Then we employ rank correlation analysis in order to compare these new rankings and the original one to test the robustness of the CIP ranking.

The scheme of the research partially replicates that of our previous work on aggregate rankings of academic journals [Aleskerov et al., 2011, Aleskerov et al., 2013, Aleskerov et al., 2014].

The text is organized as follows. In Section 2 the original formula of the CIP Index is described. In Section 3 definitions are given for two majority rule-based ranking methods (the Copeland rule and Markovian method) and for three social choice solution concepts known as tournament solutions (the uncovered set, the minimal externally stable set, and the weak top cycle). The sorting procedure based on a tournament solution is formally described in this Section. The values of correlation measures for both aggregate rankings and single-factor-based rankings are presented in Section 4. Section 5 contains formal comparison of rankings based on their correlation. Interpretation of the results and suggestions for further research are presented in Conclusion.

## 2    Competitive Industrial Performance Index

The Competitive Industrial Performance (CIP) index is a composite indicator proposed by experts of the United Nations Industrial Development Organization (UNIDO). It was first published in Industrial Development Report 2002/2003. Since then it has undergone two revisions.

The authors of the report define competitiveness as "the capacity of countries to increase their presence in international and domestic markets whilst developing industrial sectors and activities with higher value added and technological content dealing with international and domestic market shares and degree of industrialization" [UNIDO, 2013]. In its present form, the CIP index is an aggregate of eight observable variables, which represent different aspects of industrial performance. The factors are grouped into three sets or dimensions:

Dimension I. Capacity to produce and export manufactures. It is measured by

1. MVApc – manufacturing value added per capita;
2. MXpc – manufactured exports per capita;

Dimension II. Technological deepening and upgrading. It is composed of

Subdimension IIa. Industrialization intensity. It is measured by

3. MHVAsh – medium- and high-tech manufacturing value added share in in total manufacturing value added;
4. MVAsh – manufacturing value added share in total GDP;

Subdimension IIb. Manufactured Exports Quality. It is measured by

5. MHXsh – medium- and high-tech manufactured exports share in total manufactured exports;

6. MXsh – manufactured exports share in total exports;

Dimension III. World impact. It is measured by

7. ImWMVA – impact of a country on world manufacturing value added, as measured by a country's share in world MVA;

8. ImWMT – impact of a country on world manufactures trade, as measured by a country's share in world manufactured exports.

Two pairs of indicators (MVApc, MXpc and MHXsh, MXsh) are aggregated into two larger indicators by taking their arithmetic mean. The resulting CIP Index value is a product of these six factors and can be written as follows:

$$CIP = MVApc \cdot MXpc \cdot \frac{MHVAsh+MVAsh}{2} \cdot \frac{MHXsh+MXsh}{2} \cdot ImWMWA \cdot ImWMT \qquad (1)$$

A ranking is an ordered set of positions occupied by alternatives compared (in our case – countries). A rank is a number of a position. A position in an ordering can be occupied by several countries, it is said then that such countries have coinciding ranks. Positions are ordered from "best" to "worst", with their ranks increasing. In the present paper we use data provided for the year 2010 in Competitive Industrial Performance Report 2012/2013 [UNIDO, 2013]. First, countries are ranked in descending order by the values of each of eight basic indicators of UNIDO model. Then eight resulting rankings are aggregated into a single one. Countries' ranks in all rankings considered are presented in Table 6 in Appendix.

## 3      Aggregate rankings constructed by ordinal methods borrowed from social choice

Ranking of countries by values of a set of indicators is a multi-criteria evaluation problem. A common solution to a multi-criteria evaluation problem is to calculate a weighted sum of criteria values for each alternative and then rank alternatives by the value of the sum. As far as the order of alternatives is concerned, multiplying powers of criteria values is equivalent to weighted summation of their logarithms, weights being equal to powers. However, this approach has two fundamental deficiencies related to its *cardinal* nature. First, to obtain meaningful results one needs to be sure that it is theoretically possible and meaningful to perform the operation of summation and subtraction on the values of criteria or their logarithms in a given case since it is not possible generally. Second, the choice of weights (or powers) needs to be justified. The choice of weights is based on the Laplace principle, evidently. Operations in formula (1) are mathematically correct, but their results are meaningless by themselves. Only their binary comparisons make sense. Therefore we propose to apply purely *ordinal* ranking methods in order to test the robustness of the global ranking presented in UNIDO's report. We borrowed them from social choice theory since it is possible to frame any multi-criteria decision problem as a social choice problem [Arrow, Raynaud, 1986].

## 3.1 Basic notions

One of the main objectives of social choice theory is to determine what alternatives *will* be or *should* be chosen from all feasible alternatives on the basis of preferences that voters (i.e. individual participants in a collective decision-making process) have concerning these alternatives. It is possible to transfer social choice methods to a multi-criteria setting if one treats a ranking based on a certain criterion as a representation of preferences of a certain voter (or an expert). In our case, the set of rankings based on corresponding industrial performance factors is treated as a profile of preferences of eight virtual voters/experts.

Let $A$, $|A|=m$, $m \geq 3$, denote the general set of feasible alternatives; let $N$, $|N|=n$, $n \geq 2$ denote a group of experts making a collective decision by vote. A decision is a choice of certain alternatives from $A$. Preferences of a voter $i$, $i \in N$, with regard to alternatives from $A$ are revealed through pairwise comparisons of alternatives and thus are modelled by a binary relation $P_i$ on $A$, $P_i \subseteq A \times A$: if comparing an alternative $x$ with an alternative $y$ a voter $i$ prefers $x$ to $y$, then the ordered pair $(x, y)$ belongs to the relation $P_i$, $(x, y) \in P_i$; it is also said that $x$ dominates $y$ with respect to $P_i$, $xP_iy$. If a voter is unable to compare two alternatives or thinks they are of equal value, we will presume that he is indifferent regarding the choice between them, i.e. $(x, y) \notin P_i$ & $(y, x) \notin P_i$.

If chooser's preferences are known and a choice rule (a mapping of the set of binary relations on $A$ onto the set of nonempty subsets of $A$) is given, then it is possible to determine what alternatives should be the result of his choice. Thus the social choice problem can be solved if one: 1) knows individual preferences, 2) defines a binary relation $\mu$, $\mu \subseteq A \times A$ that models collective preferences (i.e. collective opinion with regard to alternatives from $A$), and 3) determines a choice rule $S(\mu, A)$: $\{\mu\} \rightarrow 2^A \setminus \emptyset$, also called a solution. Probably the most popular method to construct $\mu$ from individual preferences is to apply the majority rule. In this case, $\mu$ is called a majority (preference) relation: $x$ dominates $y$ via $\mu$ if the number of voters who prefer $x$ to $y$ is greater than the number of those who prefer $y$ to $x$, $x\mu y \Leftrightarrow |N_1| > |N_2|$, where $N_1 = \{i \in N | xP_iy\}$, $N_2 = \{i \in N | yP_ix\}$.

The choice of this particular rule of aggregation is prescribed by the social choice theory since the majority rule, and this rule only, satisfies several important normative conditions (see [Aizerman, Aleskerov, 1983]), such as independence of irrelevant alternatives, Pareto-efficiency, neutrality (equal treatment of alternatives), and anonymity (equal treatment of voters), which hold in our case as well. Moreover, in a multi-criteria setting the application of this rule allows one to obtain aggregated evaluations of alternatives without recourse to arithmetic operations on criteria, and consequently removes the problem of their theoretical justification.

We would like to test the robustness of the model with respect to change of aggregation method. Therefore we will choose weights on principle of equal treatment of factors. In the original formula six factors are treated as being of equal importance since they have the same power. Four of this factors (MVApc, MXpc, ImWMVA and ImWMT) are independent indicators. Therefore we should presume they must have the same weight. Two of this factors are arithmetic means of another two indicators, consequently all pared indicators (i.e. MVApc, MXpc and MHXsh, MXsh) are supposed to be of the same importance. Since pairs of these indicators are equal in importance with other four factors, we have to assume that the authors of the CIP index suppose that any unpaired indicator is twice as important as any paired one. We reflect this difference in importance by giving 1 vote to a virtual voter representing a paired indicator and 2 votes to a voter representing an unpaired one.

It follows from the definition that any $\mu$ is asymmetric, $(x, y) \in \mu \Rightarrow (y, x) \notin \mu$. If the following holds $x \neq y \wedge (x, y) \notin \mu \wedge (y, x) \notin \mu$, then alternatives $x$ and $y$ are tied, and both ordered pairs belong to a set of ties $\tau$, $\tau \subseteq A \times A$, $(x, y) \in \tau$ & $(y, x) \in \tau$. It is evident that a set of ties $\tau$ is an irreflexive and symmetric binary relation.

For computational purposes a majority relation $\mu$ is represented by a majority matrix $\mathbf{M} = [m_{xy}]$, defined in the following way:

$$m_{xy} = 1 \Leftrightarrow (x, y) \in \mu \text{ or } m_{xy} = 0 \Leftrightarrow (x, y) \notin \mu.$$

A matrix $\mathbf{T} = [t_{ij}]$ representing a set of ties $\tau$ is defined in the same way.

To define several choice rules we will also need the notions of the lower section, the upper section and the horizon of the alternative $x$. The lower section of an alternative $x$ is the set $L(x)$ of all alternatives dominated by $x$ via $\mu$, $L(x) = \{y | x \mu y\}$, the upper section of $x$ is the set $D(x)$ of all alternatives that dominate $x$ via $\mu$, $D(x) = \{y | y \mu x\}$, the horizon of $x$ is the set $H(x)$ of all alternatives that tie $x$, $H(x) = \{y | y \tau x\}$.

## 3.2 The Copeland rule

A majority relation quite often happens not to be a ranking itself since it is generally nontransitive. That is, a majority relation often contains cycles. For instance, there are often alternatives $x$, $y$ and $z$ such that $x \mu y$ and $y \mu z$ and $z \mu x$ (a 3-step $\mu$-cycle: $x$ is majority preferred to $y$, which is majority preferred to $z$, which is majority preferred to $x$). This result is known as the "Condorcet paradox". In order to check if majority relation in our case is transitive or not and to evaluate how nontransitive it is, we calculate the number of 3-step μ-cycles, 4-step μ-cycles and 5-step μ-cycles for our set of countries. This can be done by raising a majority matrix **M** to the power of 3, 4 and 5, correspondingly. When *k* equals 3, 4 or 5, the number of *k*-step μ-cycles $q_k$ is equal to the trace (the sum of all diagonal entries) of the matrix $\mathbf{M}^k$ divided by *k*: $q_k = \frac{\text{tr}(\mathbf{M}^k)}{k}$ [Cartwright, Gleason, 1966]. Numbers of cycles for each *k* are given in Table 1.

**Table 1. Numbers of 3-, 4- and 5-step cycles in μ**

|  | Number of cycles |
|---|---|
| 3-step cycles | 638 |
| 4-step cycles | 5928 |
| 5-step cycles | 52754 |

As we see, the Condorcet paradox occurs in our case. In order to bypass the nontransitivity problem, several ranking methods have been proposed. Probably the simplest one is the Copeland rule [Copeland, 1951]. The idea of this method is the following: the greater the number of alternatives that are worse than a given one, the better this alternative is; and it is determined through pairwise comparisons (based on a majority relation) whether a given alternative is either better or worse than another one. Alternatively, it could be put that an alternative is good if the number of alternatives that are better is small. Finally, one can combine these two principles.

Formally, the Copeland aggregate ranking is an ordering of the alternatives by their score *s*(*x*) (called the Copeland score), as given by one of the following formulae:

Version 1. $s_1(x)=|L(x)|-|D(x)|$

Version 2. $s_2(x)=|L(x)|$

Version 3. $s_3(x)=|A|-|D(x)|$

All three versions yield the same result when there are no ties. Vectors $\mathbf{s}_1$, $\mathbf{s}_2$ и $\mathbf{s}_3$ of scores, which countries obtain according to these versions, are computed by the formulae: $\mathbf{s}_2=\mathbf{M}\cdot\mathbf{a}$, $\mathbf{s}_3=(\mathbf{I}-\mathbf{M}^{\text{tr}})\cdot\mathbf{a}$, $\mathbf{s}_1= \mathbf{s}_2 + \mathbf{s}_3 - n\cdot\mathbf{a}$, where **I** and **a** denote, correspondingly, the matrix and the vector, which entries and components are all equal to 1.

Example 1. Let us consider how the second version of the Copeland rule ranks countries in the following example. Let us assume that there are $m=5$ countries, $A=\{x_1, x_2, x_3, x_4, x_5\}$, and $n=3$ factors generating three rankings. Let countries be ordered $x_1>x_2>x_3>x_4>x_5$ by the 1st factor, $x_4>x_5>x_2>x_3>x_1$ by the 2nd factor, $x_5>x_3>x_1>x_2>x_4$ by the 3d factor. The majority matrix **M** and the Copeland score (cardinality of the lower section) of a given country are presented in Table 2.

**Table 2. Majority matrix and the Copland score in Example 1**

|  | Majority matrix **M** | | | | | Cardinality of the lower section $|L(x)|$ |
|---|---|---|---|---|---|---|
|  | $x_1$ | $x_2$ | $x_3$ | $x_4$ | $x_5$ | |
| $x_1$ | 0 | 1 | 0 | 1 | 0 | 2 |
| $x_2$ | 0 | 0 | 1 | 1 | 0 | 2 |
| $x_3$ | 1 | 0 | 0 | 1 | 0 | 2 |
| $x_4$ | 0 | 0 | 0 | 0 | 1 | 1 |
| $x_5$ | 1 | 1 | 1 | 0 | 0 | 3 |

According to the second version of the Copeland rule, the aggregate ranking contains three ranks: 1) $x_5$; 2) $x_1$ - $x_2$ - $x_3$; 3) $x_4$.

## 3.3 A sorting procedure based on tournament solutions

In order to construct a ranking, we can also use solutions to the problem of optimal social choice. Let us consider the following iterative procedure. A solution concept $S(\mu, A)$ is a choice correspondence that determines a set $B_{(1)}$ of those alternatives from a set $A$ that are considered to be the best with respect to collective preferences expressed in a form of a majority relation $\mu$: $B_{(1)}=S(\mu, A)$. Alternatives from $B_{(1)}$ are of "prime quality" choices comparing with all other alternatives. Let us exclude them and repeat the sorting procedure for the set $A \backslash B_{(1)}$. Then a set $B_{(2)}=S(\mu, A \backslash B_{(1)})=S(\mu, A \backslash S(\mu, A))$ will be determined. This set contains second best choices – they are worse than alternatives from $B_{(1)}$ and better than options from $A \backslash (B_{(1)} \cup B_{(2)})$). After a finite number of selections and exclusions, all alternatives from A will be separated by classes $B_{(k)}=S(\mu, A \backslash (B_{(k-1)} \cup B_{(k-2)} \cup ... \cup B_{(2)} \cup B_{(1)}))$ according to their "quality", and these classes define the ranking we are looking for.

In this study, we use two tournament solutions: the uncovered set and the externally stable set. The first solution is based on the following idea: let us make the notion of majority preferences stronger, so it becomes always possible to choose undominated alternatives[1]. That is, when the set of undominated alternatives of μ is empty, let us select undominated alternatives of a special subset α of μ, α⊆μ. The subrelation α is defined in the following way. It is said that an alternative $x$ covers $y$, $x\alpha y$, if $x$ μ-dominates both $y$ and all alternatives, which are μ-dominated by $y$: $x\alpha y \Leftrightarrow (x\mu y \land \forall z \in A \, (y\mu z \Rightarrow x\mu z))$ [Miller, 1980]. That is, the majority of voters strongly prefer $x$ to $y$ when 1) they prefer $x$ to $y$, and 2) there is no alternative $z$, such that it is strictly less preferable than $y$ and at least as preferable as $x$. The best alternatives are those not covered (not dominated with respect to α) by any other alternatives. Their set is called the uncovered set[2] *UC*. The uncovered set is always nonempty due to the transitivity of the covering relation α.

Instead of choosing "strong" candidates as is the case with the uncovered set, it is possible to choose candidates from a "strong" group. The second solution is based on this idea of choosing from a set endowed with some "good" properties. A set *ES* is externally stable if for any alternative $x$ outside *ES* there exists an alternative $y$ in *ES* that is more preferable for the majority of voters than $x$: $\forall x \notin ES \, \exists y: y \in ES \land y\mu x$ [von Neumann, Morgenstern, 1944]. An externally stable set is minimal if none of its proper subsets is externally stable. An alternative is optimal if it belongs to at least one minimal externally stable set *MES*, therefore the tournament solution is the union of all such sets, which is likewise denoted as *MES* ([Subochev, 2008]; see also [Aleskerov, Subochev, 2013])[3]. *MES* is always nonempty.

---

[1] Due to the Condorcet paradox, the set of alternatives undominated via the majority relation itself (the so-called core) may (and almost always will) be empty.

[2] There exist alternative definitions of the covering relation and, consequently, of the uncovered set. They are listed in Aleskerov, Subochev (2013).

[3] Minimal externally stable set was introduced by Subochev (2008) as a version of another tournament solution – minimal weakly stable set (MWS) introduced by Aleskerov and Kurbanov (1999). Therefore in Subochev (2008) and in Aleskerov, Subochev (2009) this solution concept is called the second version of the minimal weakly stable set and is denoted as $MWS^{II}$. The version of the uncovered set we use here is denoted as $UC^{II}$ in the aforementioned texts.

When *UC* (or *MES*) is determined for the initial set of countries, the countries comprised by this set receive the first (best) rank. After that, these countries are excluded from the general set *A* and the procedure repeats iteratively, as it was explained in the beginning of this section.

The uncovered set and the union of minimal externally stable sets can be calculated through their matrix-vector representations given in Aleskerov, Subochev [2009; 2013]. These representations use the matrices **M** and **T** defined in Subsection 3.1.

## 3.4 The Markovian method

Finally, we would like to apply a version of a ranking called the Markovian method, since it is based on an analysis of Markov chains that model stochastic moves from vertex to vertex via arcs of a digraph representing a binary relation µ. The earliest versions of this method were proposed by Daniels [1969] and Ushakov [1971]. References to other papers can be found in Chebotarev, Shamis [1999].

To explain the method let us consider its application in the following situation. Suppose alternatives from *A* are chess-players. Only two persons can sit at a chess-board, therefore in making judgments about players' relative strength, we are compelled to rely upon results of binary comparisons, i.e. separate games. Our aim is to rank players according to their strength. Since it is not possible with a single game, we organize a tournament.

Before the tournament starts we separate patently stronger players from the weaker ones by assigning each player to a certain league, a subgroup of players who are relatively equal in their strength. To make the assignments, we use the sorting procedure described in the previous subsection. The tournament solution that is used for the selection of the strongest players is the weak top cycle *WTC* [Ward, 1961; Schwartz, 1970, 1972, 1977; Good, 1971; Smith, 1973]. It is defined in the following way. A set *WTC* is called the weak top cycle if 1) any alternative in *WTC* µ-dominates any alternative outside *WTC*: $\forall$ $x \notin WTC$, $y \in WTC \Rightarrow y\mu x$, and 2) none of its proper subsets satisfies this property.

The relative strength of players assigned to different leagues is determined by a binary relation μ, therefore in order to rank all players all we need to know is how to rank players of the same league. Each league receives a chess-board. Since there is only one chess-board per league, the games of a league form a sequence in time.

Players who participate in a game are chosen in the following way: a player who has been declared a (current) winner in the previous game remains at the board, her rival is randomly chosen from the rest of the players, among whom the loser of the previous game is also present. In a given league, all probabilities of being chosen are equal. If a game ends in a draw, the previous winner, nevertheless, loses her title and it passes to her rival. Therefore, despite ties being allowed, there is a single winner in each game. It is evident that the strength of a player can be measured by counting a relative number of games where he has been declared a winner (i.e. the number of his wins divided by the total number of games in a tournament).

In order to start a tournament we need to decide who is declared a winner in a fictitious "zero-game". However, the longer a tournament goes (i.e. the greater the number of tournament games is), the smaller is the influence of this decision on the relative number of wins of any player. In the limit when the number of games tends to infinity relative numbers of wins are completely independent of who had been given "the crown" before the tournament started.

Instead of calculating the limit of the relative number of wins, one can find the limit of the probability a player will be declared a winner in the last game of the tournament since these values are equal. We can count the probability and its limit using matrices **M** and **T** defined above.

Suppose we somehow know the relative strength of players in each pair of them. Also, suppose this strength is constant over time and is represented by binary relations μ and τ. Therefore, if we know μ and the names of the players who are sitting at the chess-board, we can predict the result of the game: the victory of $x$ (if $x\mu y$), the victory of $y$ (if $y\mu x$) or a draw (if $x\tau y$).

Let $\mathbf{p}^{(k)}$ denote a vector, $i$-th component $p_i^{(k)}$ of which is the probability a player number $i$

is declared the winner of a game number *k*. Two mutually exclusive situations are possible. The first case – the player number *i* is declared the winner in both the previous game (game number *k*-1) and the current game. She can be declared the winner in the game number *k*, if and only if her rival (who has been chosen by lot) belongs to the lower section of *i*. The probability that the *i*-th player was declared the winner in the game number *k*-1 is $p_i^{(k-1)}$, the probability of her rival being in *L(i)* equals $\frac{s_2(i)}{m-1}$, where $s_2(i)$ is the Copeland score (the 2$^{nd}$ version), $s_2(x)=|L(x)|$. Thus, the probability of the *i*-th player being declared the winner in game number *k* is $p_i^{(k-1)} \cdot \frac{s_2(i)}{m-1}$.

The second case – the player number *i* is declared the winner in the current game, but not in the previous one. He can be declared the winner in game number *k*, if and only if 1) he has been chosen by lot as a rival to the winner in the game number *k*-1, the probability of which equals $\frac{1}{m-1}$; and 2) if the (*k*-1)-th winner is in the lower section or in the horizon of the *i*-th player, a probability of which equals $\sum_{j=1}^{m}(m_{ij} + t_{ij}) \cdot p_j^{(k-1)}$.[4] Thus the probability $p_i^{(k)}$ can be determined from the following equation:

$$p_i^{(k)} = p_i^{(k-1)} \cdot \frac{s_2(i)}{m-1} + \frac{1}{m-1} \cdot \sum_{j=1}^{m}(m_{ij} + t_{ij}) \cdot p_j^{(k-1)} \qquad (2)$$

Formula (2) can be rewritten in a matrix-vector form as:

$$\mathbf{p}^{(k)} = \mathbf{W} \cdot \mathbf{p}^{(k-1)} = \frac{1}{m-1} \cdot (\mathbf{M} + \mathbf{T} + \mathbf{S}) \cdot \mathbf{p}^{(k-1)} \qquad (3)$$

The matrix $\mathbf{S}=[s_{ij}]$ is defined thus: $s_{ii}=s_2(i)$ and $s_{ij}=0$ when $i \neq j$.

Consequently, passing the title of the current winner from player to player is a Markovian process with the transition matrix **W**.

We are interested in vector $\mathbf{p}=\lim_{k \to \infty} \mathbf{p}^{(k)}$. It is not hard to prove that no matter what the initial conditions are (i.e. what the value of $\mathbf{p}^{(0)}$ is), the limit vector is an eigenvector of the matrix **W** corresponding to the eigenvalue λ=1 (see, for instance, Laslier [1997]). Therefore **p** is determined by solving the system of linear equations **W·p=p**. To rank players in a league, one needs to order them by decreasing values of $p_i$. Since we have pre-sorted players using *WTC*, none of the components $p_i$ is equal to zero [Laslier, 1997].

---

[4] Here notations *m*, $m_{ij}$, $t_{ij}$ are those introduced in Subsection 3.1.

Ranks of the countries in the six aggregate rankings are given in Table 6 in Appendix.

## 4    Correlations

The number of the alternative's position in a ranking is a rank variable. Therefore, to evaluate the (in)consistency of two rankings, one needs to apply ranking measures of correlation. In this paper, we use two related but not identical measures based on the Kendall distance: the Kendall rank correlation index $\tau_b$ [Kendall, 1938] and the share of coinciding pairs *r*.

To remind the reader what the Kendall distance is, let us consider a pair of countries and compare their positions in two rankings. If a country is placed above the second one in the first ranking, but at the same time it is placed below the other one in the second ranking, then this pair of countries counts as an inversion. The Kendall distance between two rankings is the number of inversions $N_-$ (a number of unordered pairs of objects ranked inversely in two ranking). Correspondingly, the greater the number of inversions is, the farther apart (i.e. the more disparate) the rankings are. The Kendall rank correlation coefficient $\tau_b$ depends on the Kendall distance in the following way:

$$\tau_b = \frac{N_+ - N_-}{\sqrt{(N-n_1)\cdot(N-n_2)}} \qquad (3)$$

Here $N_+$ is the number of coinciding pairs, which are not ties, i.e. such country pairs, where one country is placed above the second one in both rankings; $n_1$ is the number of pairs, where both countries have the same rank in the first ranking; $n_2$, correspondingly, is the number of pairs, where both countries have the same rank in the second ranking. Obviously, $N_+ + N_- = N - n_1 - n_2 + N_0$, where $N_0$ is the number of pairs tied in both rankings.

The share of coinciding pairs *r* is a percentage of pairs ranked in the same way in both rankings, $r = 100 \cdot \frac{N_+ + N_0}{N}$. This measure has a simple probabilistic interpretation. If we know that alternative *x* is ranked above alternative *y* in ranking $R_1$ and guess that in ranking $R_2$ they are placed in the same order, then *r* is the probability of us being correct. When *r*=50%, probability of being right equals probability of being wrong, which means two rankings do not correlate.

The main difference between $\tau_b$ and *r* is that the latter "punishes" rankings containing too many ties, while the former does not. Values of $\tau_b$ and *r* for the eight factor-based and aggregate rankings are given in Table 3.

**Table 3. Values of correlation measures**

| | MVApc | MXpc | MHVAsh | MVAsh | MHXsh | MXsh | ImWMVA | ImWMT | the CIP index | Copeland (1) | Copeland (2) | Copeland (3) | UC | MES | Markovian |
|---|---|---|---|---|---|---|---|---|---|---|---|---|---|---|---|
| | | | | | | | Kendall's $\tau_b$ | | | | | | | | |
| MVApc | 1,000 | 0,767 | 0,476 | 0,318 | 0,465 | 0,365 | 0,510 | 0,553 | 0,715 | 0,718 | 0,715 | 0,723 | 0,714 | 0,691 | 0,714 |
| MXpc | 0,767 | 1,000 | 0,487 | 0,289 | 0,466 | 0,421 | 0,440 | 0,576 | 0,704 | 0,716 | 0,716 | 0,716 | 0,706 | 0,689 | 0,709 |
| MHVAsh | 0,476 | 0,487 | 1,000 | 0,319 | 0,471 | 0,399 | 0,517 | 0,578 | 0,595 | 0,637 | 0,633 | 0,643 | 0,654 | 0,635 | 0,633 |
| MVAsh | 0,318 | 0,289 | 0,319 | 1,000 | 0,319 | 0,381 | 0,436 | 0,422 | 0,440 | 0,456 | 0,455 | 0,458 | 0,471 | 0,476 | 0,448 |
| MHXsh | 0,465 | 0,466 | 0,471 | 0,319 | 1,000 | 0,354 | 0,422 | 0,470 | 0,529 | 0,559 | 0,563 | 0,556 | 0,576 | 0,571 | 0,556 |
| MXsh | 0,365 | 0,421 | 0,399 | 0,381 | 0,354 | 1,000 | 0,289 | 0,370 | 0,430 | 0,476 | 0,472 | 0,482 | 0,492 | 0,472 | 0,485 |
| ImWMVA | 0,510 | 0,440 | 0,517 | 0,436 | 0,422 | 0,289 | 1,000 | 0,808 | 0,732 | 0,701 | 0,703 | 0,701 | 0,717 | 0,720 | 0,679 |
| ImWMT | 0,553 | 0,576 | 0,578 | 0,422 | 0,470 | 0,370 | 0,808 | 1,000 | 0,833 | 0,801 | 0,805 | 0,798 | 0,808 | 0,801 | 0,774 |
| CIP | 0,715 | 0,704 | 0,595 | 0,440 | 0,529 | 0,430 | 0,732 | 0,833 | 1,000 | 0,930 | 0,926 | 0,925 | 0,907 | 0,877 | 0,888 |
| Cop. (1) | 0,718 | 0,716 | 0,637 | 0,456 | 0,559 | 0,476 | 0,701 | 0,801 | 0,930 | 1,000 | 0,979 | 0,982 | 0,937 | 0,897 | 0,921 |
| Cop. (2) | 0,715 | 0,716 | 0,633 | 0,455 | 0,563 | 0,472 | 0,703 | 0,805 | 0,926 | 0,979 | 1,000 | 0,959 | 0,936 | 0,899 | 0,905 |
| Cop. (3) | 0,723 | 0,716 | 0,643 | 0,458 | 0,556 | 0,482 | 0,701 | 0,798 | 0,925 | 0,982 | 0,959 | 1,000 | 0,935 | 0,896 | 0,933 |
| UC | 0,714 | 0,706 | 0,654 | 0,471 | 0,576 | 0,492 | 0,717 | 0,808 | 0,907 | 0,937 | 0,936 | 0,935 | 1,000 | 0,915 | 0,913 |
| MES | 0,691 | 0,689 | 0,635 | 0,476 | 0,571 | 0,472 | 0,720 | 0,801 | 0,877 | 0,897 | 0,899 | 0,896 | 0,915 | 1,000 | 0,878 |
| Markovian | 0,714 | 0,709 | 0,633 | 0,448 | 0,556 | 0,485 | 0,679 | 0,774 | 0,888 | 0,921 | 0,905 | 0,933 | 0,913 | 0,878 | 1,000 |
| | | | | | | | Percentage of coinciding pairs (*r*) | | | | | | | | |
| MVApc | 100 | 88,36 | 73,80 | 65,89 | 73,27 | 68,24 | 74,89 | 77,13 | 85,66 | 85,77 | 85,32 | 85,59 | 81,77 | 78,78 | 85,72 |
| MXpc | 88,36 | 100 | 74,32 | 64,46 | 73,30 | 71,02 | 71,43 | 78,25 | 85,11 | 85,65 | 85,34 | 85,23 | 81,38 | 78,70 | 85,44 |
| MHVAsh | 73,80 | 74,32 | 100 | 65,91 | 73,55 | 69,93 | 75,23 | 78,33 | 79,65 | 81,68 | 81,22 | 81,57 | 78,84 | 76,10 | 81,65 |
| MVAsh | 65,89 | 64,46 | 65,91 | 100 | 65,96 | 69,02 | 71,23 | 70,55 | 71,92 | 72,67 | 72,33 | 72,40 | 69,96 | 68,49 | 72,37 |
| MHXsh | 73,27 | 73,30 | 73,55 | 65,96 | 100 | 67,68 | 70,53 | 72,98 | 76,36 | 77,82 | 77,73 | 77,27 | 75,10 | 73,03 | 77,80 |
| MXsh | 68,24 | 71,02 | 69,93 | 69,02 | 67,68 | 100 | 63,89 | 68,00 | 71,42 | 73,63 | 73,19 | 73,60 | 71,01 | 68,33 | 74,24 |
| ImWMVA | 74,89 | 71,43 | 75,23 | 71,23 | 70,53 | 63,89 | 100 | 89,46 | 85,86 | 84,28 | 84,08 | 83,89 | 81,49 | 79,73 | 83,34 |
| ImWMT | 77,13 | 78,25 | 78,33 | 70,55 | 72,98 | 68,00 | 89,46 | 100 | 90,98 | 89,29 | 89,22 | 88,77 | 85,97 | 83,62 | 88,13 |
| CIP | 85,66 | 85,11 | 79,65 | 71,92 | 76,36 | 71,42 | 85,86 | 90,98 | 100 | 96,24 | 95,75 | 95,56 | 91,14 | 87,68 | 94,34 |
| Cop. (1) | 85,77 | 85,65 | 81,68 | 72,67 | 77,82 | 73,63 | 84,28 | 89,29 | 96,24 | 100 | 98,40 | 98,40 | 92,65 | 88,66 | 95,91 |
| Cop. (2) | 85,32 | 85,34 | 81,22 | 72,33 | 77,73 | 73,19 | 84,08 | 89,22 | 95,75 | 98,40 | 100 | 96,95 | 92,59 | 88,68 | 94,80 |
| Cop. (3) | 85,59 | 85,23 | 81,57 | 72,40 | 77,27 | 73,60 | 83,89 | 88,77 | 95,56 | 98,40 | 96,95 | 100 | 92,38 | 88,62 | 96,06 |
| UC | 81,77 | 81,38 | 78,84 | 69,96 | 75,10 | 71,01 | 81,49 | 85,97 | 91,14 | 92,65 | 92,59 | 92,38 | 100 | 89,08 | 91,45 |
| MES | 78,78 | 78,70 | 76,10 | 68,49 | 73,03 | 68,33 | 79,73 | 83,62 | 87,68 | 88,66 | 88,68 | 88,62 | 89,08 | 100 | 87,72 |
| Markovian | 85,72 | 85,44 | 81,65 | 72,37 | 77,80 | 74,24 | 83,34 | 88,13 | 94,34 | 95,91 | 94,80 | 96,06 | 91,45 | 87,72 | 100 |

All eight basic single-indicator-based rankings correlate positively with respect to both measures ($\tau_b$>0; *r*>50%). Their correlation is moderately strong ($\tau_b$>0,3; *r*>65%) in most cases. It is very strong ($\tau_b$>0,75; $r$>85%) in two cases: {ImWMT, ImWMVA}, {MVApc, MXpc}. This is because national manufacturing value added and manufactured exports correlate strongly.

Direct observations of values in Tables 3 also confirm natural expectations: all aggregate rankings, both old one and new ones, are better correlated with the set of eight single-indicator-based rankings than the latter with each other.

Original CIP ranking correlate strongly and positively with all new aggregate rankings, the lowest level of contradictions being 3,76% (with the 1$^{st}$ version of the Copeland rule), the highest – 12,32% (with the ranking based on *MES*). The pair {CIP, *MES*} demonstrated the lowest correlation among all pairs of all aggregate rankings according to both measures. Therefore we can use values of $\tau_b$ and $r$ for this pair in order to evaluate robustness of CIP. We may conclude that strong ($\tau_b$>0,75; $r$>85%) correlation of these two ranking support the claim that the CIP ranking is robust.

One can observe that values of $r$ for pairs of aggregate rankings vary greater than their values of $\tau_b$. This difference between two measures can be explained thus: the scales of rankings produced by sorting contain too few grades as compared to scales of other rankings, consequently rankings based on *UC* and *MES* contain significantly more ties than other rankings. As a result values of $r$ for pairs containing either of this two rankings are lower, since this measure (unlike $\tau_b$) "punishes" rankings containing too many ties: being a tie in a ranking based on *UC* or *MES*, a pair most probably will not be a tie in another ranking and so it will not contribute to the numerator of $r$, while $r$'s denominator remains constant across all pairs.

## 5 Formal comparison of rankings

Let us employ the same method of binary multi-criteria comparisons to analyze rankings more formally. The problem of aggregation can be reformulated as a choice of a single object representing a given group of objects. In our case we need to choose a ranking that will represent the set of eight single-indicator-based rankings $\{P_i\}$, $i=1\div8$. We have fifteen candidates: seven aggregate rankings and eight initial rankings. Let us use the same idea of binary multi-criteria comparisons and majority relations in order to determine the best representations.

Let us say that ranking $R_1$ represents single-indicator-based ranking $P_i$ better than ranking $R_2$ does if $R_1$ is better correlated with $P_i$ than $R_2$. If $P_i$ represents preferences of voter $i$ then we may suppose that $R_1$ represents $i$'s preferences better than $R_2$ does, so voter $i$ will most likely vote for $R_1$ against $R_2$, when they are compared. Then $R_1$ should be considered a better representative for the set of rankings $\{P_i\}$ than $R_2$ if $R_1$ is better correlated with (is closer to) a (weighted) majority of rankings from this set than $R_2$ is. Let us remind a reader that weight $v_i$ (the number of votes that voter $i$ has) reflects relative importance attributed to the corresponding aggregated variable $i$. In our case the vector of weights/votes is $\mathbf{v}=(2, 2, 1, 1, 1, 1, 2, 2)$.

Each ranking $R$ is characterized by 8-component vector $\mathbf{c}(R)$, its $i$-th component being the value of a given correlation measure for this ranking and corresponding single-indicator-based ranking $P_i$: either $c_i(R) = \tau_b(R, P_i)$ or $c_i(R) = r(R, P_i)$. We perform binary comparisons of vectors $\mathbf{c}(R)$ and define a majority relation on the set of twelve rankings in the following way: $R_1\, \mu\, R_2 \Leftrightarrow V_1 > V_2$, where $V_1 = \sum_{\{i|c_i(R_1)>c_i(R_2)\}} v_i$, $V_2 = \sum_{\{i|c_i(R_2)>c_i(R_1)\}} v_i$.

Table 4 contains results of binary comparisons based on measures $\tau_b$ and $r$. The first number in a cell equals 1 if the ranking of the row correlates with eight single-factor rankings better than the ranking of the column with respect to a given measure of correlation. It equals 0 otherwise, that is the first numbers are majority matrices' entries. The second number (in brackets) is a number of those initial rankings that are closer to the ranking of the row than to the ranking of the column with respect to a given measure of correlation.

**Table 4. Binary comparisons of rankings (majority matrices and numbers of "wins")**

| | MVApc | MXpc | MHVAsh | MVAsh | MHXsh | MXsh | ImWMVA | ImWMT | the CIP index | Copeland (1) | Copeland (2) | Copeland (3) | UC | MES | Markovian |
|---|---|---|---|---|---|---|---|---|---|---|---|---|---|---|---|
| Kendall's $\tau_b$ | | | | | | | | | | | | | | | |

|         | | | | | | | | | | | | | | |
|---------|---|---|---|---|---|---|---|---|---|---|---|---|---|---|
| MVApc   | 0(0)  | 1(11) | 1(10) | 1(10) | 1(8)  | 1(7)  | 0(5)  | 0(3)  | 0(1)  | 0(1)  | 0(1)  | 0(1)  | 0(1)  | 0(1)  |
| MXpc    | 0(1)  | 0(0)  | 0(4)  | 0(5)  | 0(2)  | 0(1)  | 0(2)  | 0(2)  | 0(1)  | 0(1)  | 0(1)  | 0(1)  | 0(1)  | 0(1)  |
| MHVAsh  | 0(2)  | 1(8)  | 0(0)  | 1(10) | 0(2)  | 0(2)  | 0(4)  | 0(1)  | 0(1)  | 0(1)  | 0(1)  | 0(1)  | 0(1)  | 0(1)  |
| MVAsh   | 0(2)  | 1(7)  | 0(2)  | 0(0)  | 0(2)  | 0(2)  | 0(1)  | 0(1)  | 0(1)  | 0(1)  | 0(1)  | 0(1)  | 0(1)  | 0(1)  |
| MHXsh   | 0(4)  | 1(10) | 1(10) | 1(10) | 0(0)  | 0(5)  | 0(6)  | 0(4)  | 0(4)  | 0(4)  | 0(4)  | 0(4)  | 0(4)  | 0(4)  |
| MXsh    | 0(5)  | 1(11) | 1(10) | 1(10) | 1(7)  | 0(0)  | 0(6)  | 0(5)  | 0(4)  | 0(4)  | 0(4)  | 0(4)  | 0(4)  | 0(4)  |
| ImWMVA  | 1(7)  | 1(10) | 1(8)  | 1(11) | 0(6)  | 0(6)  | 0(0)  | 0(3)  | 0(2)  | 0(4)  | 0(4)  | 0(4)  | 0(4)  | 0(4)  |
| ImWMT   | 1(9)  | 1(10) | 1(11) | 1(11) | 1(8)  | 1(7)  | 1(9)  | 0(0)  | 0(4)  | 0(4)  | 0(4)  | 0(4)  | 0(4)  | 0(4)  |
| CIP     | 1(11) | 1(11) | 1(11) | 1(11) | 1(8)  | 1(8)  | 1(10) | 1(8)  | 0(0)  | 0(4)  | 0(4)  | 0(4)  | 0(6)  | 1(8)  | 0(6)  |
| Cop. (1)| 1(11) | 1(11) | 1(11) | 1(11) | 1(8)  | 1(8)  | 1(8)  | 1(8)  | 1(8)  | 0(0)  | 1(7)  | 1(7)  | 0(4)  | 0(6)  | 1(11) |
| Cop. (2)| 1(11) | 1(11) | 1(11) | 1(11) | 1(8)  | 1(8)  | 1(8)  | 1(8)  | 1(8)  | 0(5)  | 0(0)  | 0(5)  | 0(4)  | 0(6)  | 1(10) |
| Cop. (3)| 1(11) | 1(11) | 1(11) | 1(11) | 1(8)  | 1(8)  | 1(8)  | 1(8)  | 1(8)  | 0(5)  | 1(7)  | 0(0)  | 0(4)  | 0(6)  | 1(10) |
| *UC*    | 1(11) | 1(11) | 1(11) | 1(11) | 1(8)  | 1(8)  | 1(8)  | 1(8)  | 0(6)  | 1(8)  | 1(8)  | 1(8)  | 0(0)  | 1(9)  | 1(8)  |
| *MES*   | 1(11) | 1(11) | 1(11) | 1(11) | 1(8)  | 1(8)  | 1(8)  | 1(8)  | 0(4)  | 0(6)  | 0(6)  | 0(6)  | 0(3)  | 0(0)  | 1(7)  |
| Markovian | 1(11) | 1(11) | 1(11) | 1(11) | 1(8) | 1(8) | 1(8) | 1(8) | 0(6) | 0(1) | 0(2) | 0(2) | 0(4) | 0(5) | 0(0) |
| **Percentage of coinciding pairs (*r*)** | | | | | | | | | | | | | | |
| MVApc   | 0(0)  | 1(11) | 1(10) | 1(10) | 1(8)  | 1(7)  | 0(5)  | 0(3)  | 0(1)  | 0(1)  | 0(1)  | 0(1)  | 0(1)  | 0(3)  | 0(1)  |
| MXpc    | 0(1)  | 0(0)  | 0(4)  | 0(5)  | 0(2)  | 0(1)  | 0(2)  | 0(2)  | 0(1)  | 0(1)  | 0(1)  | 0(1)  | 0(1)  | 0(2)  | 0(1)  |
| MHVAsh  | 0(2)  | 1(8)  | 0(0)  | 1(10) | 0(2)  | 0(2)  | 0(4)  | 0(1)  | 0(1)  | 0(1)  | 0(1)  | 0(1)  | 0(1)  | 0(1)  | 0(1)  |
| MVAsh   | 0(2)  | 1(7)  | 0(2)  | 0(0)  | 0(2)  | 0(2)  | 0(1)  | 0(1)  | 0(1)  | 0(1)  | 0(1)  | 0(1)  | 0(1)  | 0(2)  | 0(1)  |
| MHXsh   | 0(4)  | 1(10) | 1(10) | 1(10) | 0(0)  | 0(5)  | 0(6)  | 0(6)  | 0(4)  | 0(4)  | 0(4)  | 0(4)  | 0(4)  | 0(5)  | 0(4)  |
| MXsh    | 0(5)  | 1(11) | 1(10) | 1(10) | 1(7)  | 0(0)  | 0(6)  | 0(6)  | 0(4)  | 0(4)  | 0(4)  | 0(4)  | 0(5)  | 0(6)  | 0(4)  |
| ImWMVA  | 1(7)  | 1(10) | 1(8)  | 1(11) | 0(6)  | 0(6)  | 0(0)  | 0(3)  | 0(2)  | 0(4)  | 0(4)  | 0(4)  | 0(5)  | 0(5)  | 0(4)  |
| ImWMT   | 1(9)  | 1(10) | 1(11) | 1(11) | 0(6)  | 0(6)  | 1(9)  | 0(0)  | 0(4)  | 0(4)  | 0(4)  | 0(4)  | 0(5)  | 0(6)  | 0(4)  |
| CIP     | 1(11) | 1(11) | 1(11) | 1(11) | 1(8)  | 1(8)  | 1(10) | 1(8)  | 0(0)  | 0(4)  | 0(6)  | 0(6)  | 1(12) | 1(12) | 0(4)  |
| Cop. (1)| 1(11) | 1(11) | 1(11) | 1(11) | 1(8)  | 1(8)  | 1(8)  | 1(8)  | 1(8)  | 0(0)  | 1(12) | 1(12) | 1(12) | 1(12) | 1(11) |
| Cop. (2)| 1(11) | 1(11) | 1(11) | 1(11) | 1(8)  | 1(8)  | 1(8)  | 1(8)  | 0(6)  | 0(0)  | 0(0)  | 1(7)  | 1(12) | 1(12) | 0(4)  |
| Cop. (3)| 1(11) | 1(11) | 1(11) | 1(11) | 1(8)  | 1(8)  | 1(8)  | 1(8)  | 0(6)  | 0(0)  | 0(5)  | 0(0)  | 1(12) | 1(12) | 0(5)  |
| *UC*    | 1(11) | 1(11) | 1(11) | 1(11) | 1(8)  | 1(7)  | 1(7)  | 1(7)  | 0(0)  | 0(0)  | 0(0)  | 0(0)  | 0(0)  | 1(12) | 0(0)  |
| *MES*   | 1(9)  | 1(10) | 1(11) | 1(10) | 1(7)  | 0(6)  | 1(7)  | 0(6)  | 0(0)  | 0(0)  | 0(0)  | 0(0)  | 0(0)  | 0(0)  | 0(0)  |
| Markovian | 1(11) | 1(11) | 1(11) | 1(11) | 1(8) | 1(8) | 1(8) | 1(8) | 1(8) | 0(1) | 1(8) | 1(7) | 1(12) | 1(12) | 0(0) |

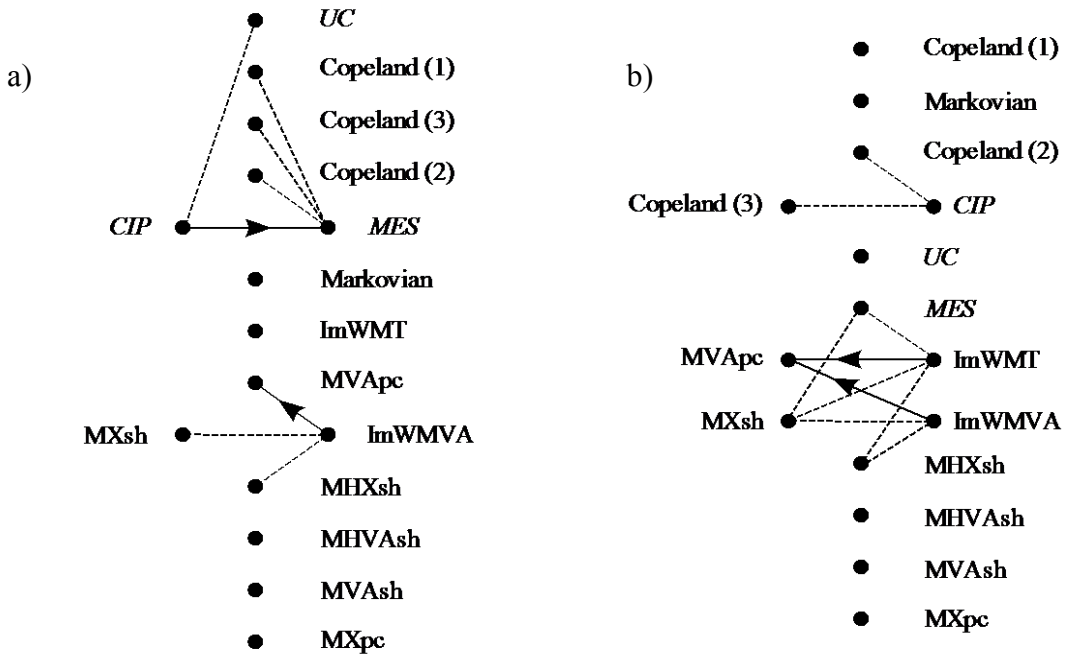

Figure 1. Ordering of rankings according to $\tau_b$ (a) and *r* (b).

A binary relation (or its matrix) can be represented by a digraph. Vertices represent alternatives, arcs (links with arrows) represent ordered pairs: the alternative, which is represented by arc's starting point, dominates (via relation represented by the digraph) the alternative, which is represented by arc's ending point. Digraphs representing matrices in Table 4 are depicted on Figure 1. By convention, if a pair of vertices is not connected it means that the arc stars at the higher vertex and goes down. A line without arrow indicates a tie.

In both cases µ is a strict partial order but not a weak order and, consequently it is not a ranking itself. We need again somehow to mend µ in order to get a ranking. First one may note that in both cases µ is very close (with respect to the Kendal distance) to a linear order (i.e. a ranking discriminating all alternatives). Therefore we can represent µ by a closest linear order. In the first case, when ranking are compared by $\tau_b$, the linear order at a minimal distance from µ is unique. In the second case, there are six closest linear order which differ only with respect to how they order the triplet {Copeland 2, Copeland 3, CIP} and the pair {*MES*, ImWMT}. In both cases the Kendall distance from µ to closest linear orders equals 0 (i.e. there are no inversions). We may unite six linear orders in a weak order assigning rank 3 to all alternatives from the triplet and rank 5 to *MES* and ImWMT. Final rankings of ranking are presented in Table 5.

**Table 5. Two rankings of rankings**

| Rank | Ordered by | |
|---|---|---|
| | $\tau_b$ | $r$ |
| 1 | *UC* | Copeland (1) |
| 2 | Copeland (1) | Markovian |
| 3 | Copeland (3) | CIP, Copeland (2), Copeland (3) |
| 4 | Copeland (2) | |
| 5 | CIP | |
| 6 | *MES* | *UC* |
| 7 | Markovian | ImWMT, *MES* |
| 8 | ImWMT | |
| 9 | ImWMVA | ImWMVA |
| 10 | MVApc | MVApc |
| 11 | MXsh | MXsh |
| 12 | MHXsh | MHXsh |
| 13 | MHVAsh | MHVAsh |
| 14 | MVAsh | MVAsh |
| 15 | MXpc | MXpc |

The following observations can be made. In all cases aggregate rankings represent the set of single-factor based rankings better than any one of latter do. Therefore replacing eight single-factor rankings by aggregate rankings is justified.

The ranking based Copeland 1$^{st}$ in both cases perform better than the CIP, but the former correlates with the latter higher than any of other aggregate ranking does.

# 6      Conclusion

The Competitive Industrial Performance index is an aggregate of eight observable variables. Its aggregation formula is semi-ordinal. It is cardinal in its form but it is derived from a purely ordinal proposition: the value of an aggregate index should be a strongly increasing function of each of its factors. Only binary comparisons of these values (and not the values themselves or values of their differences or fraction) are meaningful.

Therefore it was interesting for us to test the robustness of the final ranking by replacing the original aggregation formula by purely ordinal methods. We propose to consider aggregation as a multicriteria decision problem and to employ ordinal ranking methods borrowed from social choice to solve it. In this paper we apply two direct ranking methods based on majority rule (the Copeland rule and the Markovian method) and a multistage procedure of selection and exclusion of the best alternatives, as determined by a majority rule-based social choice solution concept (tournament solution), such as the uncovered set and the minimal externally stable set.

The Markovian ranking is characterized by high level of discrimination - it separates all 135 countries. The sorting by uncovered set and by the minimal externally stable set produced a rough division of countries into large groups - both rankings contain only 23 ranks. Intuitively, these "rough" orderings seem to be more attractive as representations of relevant differences in industrial competitiveness of nations. The ability to produce such "rough" rankings can be considered as a strength of the approach proposed.

We use the same method of binary comparisons based on majority rule to analyse rank correlations. The correlation analysis has shown that aggregate rankings are better representations for the set single-factor rankings than any one of the set. Therefore, replacing single-factor rankings by an aggregate ranking is justified. Though the high level of correlations of all aggregate rankings confirms, apparently, that the original version based on the CIP index is robust, it has also been demonstrated that some of the new aggregate rankings represent the set of criteria better.

The overall conclusion would be the following. Given the large number of different aggregation models and methods and high uncertainty concerning values of their parameters, it seems that much deeper theoretical work is needed to clarify what the national competitiveness really is and how we should measure it.

# Appendix

**Table 6. Ranks of countries in single-factor-based and aggregate rankings (countries are ordered as in the CIP ranking)**

| | MVApc | MXpc | MHVAsh | MVAsh | MHXsh | MXsh | ImWMVA | ImWMT | the CIP index | the Copeland r. (1) | the Copeland r. (2) | the Copeland r. (3) | sorting by *UC* | sorting by *MES* | Markovian |
|---|---|---|---|---|---|---|---|---|---|---|---|---|---|---|---|
| **Number of positions in a ranking** | 135 | 135 | 132 | 133 | 135 | 133 | 99 | 99 | 126 | 117 | 89 | 80 | 23 | 23 | 135 |
| Japan | 2 | 28 | 6 | 21 | 2 | 14 | 3 | 4 | 1 | 1 | 1 | 1 | 1 | 1 | 1 |
| Germany | 11 | 9 | 4 | 29 | 8 | 33 | 4 | 2 | 2 | 5 | 4 | 4 | 2 | 2 | 5 |
| USA | 8 | 39 | 9 | 55 | 15 | 52 | 1 | 3 | 3 | 8 | 7 | 6 | 1 | 1 | 8 |
| South Korea | 10 | 16 | 8 | 7 | 6 | 1 | 5 | 6 | 4 | 3 | 2 | 3 | 1 | 1 | 4 |
| Taiwan (China) | 7 | 15 | 3 | 6 | 7 | 4 | 10 | 11 | 5 | 2 | 2 | 2 | 1 | 1 | 2 |
| Singapore | 1 | 1 | 1 | 14 | 10 | 23 | 28 | 18 | 6 | 4 | 3 | 3 | 1 | 1 | 3 |
| China | 54 | 54 | 24 | 2 | 20 | 2 | 2 | 1 | 7 | 9 | 10 | 4 | 2 | 1 | 6 |
| Switzerland | 3 | 5 | 35 | 30 | 9 | 15 | 24 | 16 | 8 | 6 | 5 | 5 | 2 | 2 | 7 |
| Belgium | 14 | 2 | 19 | 52 | 31 | 30 | 26 | 9 | 9 | 11 | 8 | 8 | 2 | 3 | 12 |
| France | 21 | 23 | 13 | 73 | 14 | 27 | 7 | 5 | 10 | 12 | 9 | 8 | 3 | 3 | 11 |
| Italy | 22 | 24 | 27 | 53 | 33 | 14 | 8 | 7 | 11 | 16 | 14 | 10 | 4 | 4 | 18 |
| Netherlands | 17 | 6 | 25 | 71 | 30 | 56 | 23 | 8 | 12 | 13 | 11 | 9 | 3 | 3 | 21 |
| Sweden | 5 | 7 | 10 | 25 | 25 | 24 | 20 | 21 | 13 | 7 | 6 | 6 | 2 | 2 | 9 |
| UK | 19 | 31 | 20 | 83 | 17 | 47 | 6 | 10 | 14 | 17 | 15 | 11 | 4 | 4 | 14 |
| Ireland | 6 | 4 | 2 | 15 | 34 | 13 | 31 | 26 | 15 | 10 | 7 | 7 | 2 | 2 | 10 |
| Austria | 9 | 8 | 22 | 31 | 22 | 32 | 25 | 24 | 16 | 14 | 12 | 9 | 3 | 3 | 22 |
| Canada | 20 | 26 | 31 | 78 | 29 | 77 | 13 | 13 | 17 | 23 | 18 | 15 | 5 | 5 | 26 |
| Finland | 4 | 11 | 14 | 13 | 40 | 17 | 29 | 32 | 18 | 15 | 13 | 10 | 4 | 3 | 16 |
| Spain | 30 | 32 | 36 | 75 | 26 | 40 | 14 | 14 | 19 | 25 | 18 | 17 | 5 | 5 | 27 |
| Czech Republic | 27 | 12 | 17 | 8 | 11 | 18 | 38 | 25 | 20 | 18 | 15 | 12 | 4 | 4 | 15 |
| Malaysia | 34 | 27 | 21 | 10 | 16 | 41 | 27 | 17 | 21 | 22 | 17 | 14 | 5 | 5 | 17 |
| Mexico | 43 | 44 | 28 | 45 | 4 | 46 | 12 | 12 | 22 | 26 | 19 | 16 | 5 | 5 | 25 |
| Thailand | 40 | 40 | 11 | 1 | 19 | 39 | 19 | 19 | 23 | 20 | 16 | 12 | 2 | 1 | 13 |
| Denmark | 13 | 10 | 41 | 72 | 37 | 60 | 39 | 31 | 24 | 29 | 22 | 18 | 5 | 4 | 28 |
| Poland | 33 | 36 | 33 | 17 | 24 | 28 | 22 | 22 | 25 | 28 | 20 | 17 | 6 | 5 | 29 |
| Israel | 18 | 21 | 5 | 60 | 28 | 3 | 37 | 34 | 26 | 19 | 16 | 11 | 3 | 3 | 24 |
| Slovakia | 25 | 13 | 18 | 9 | 13 | 6 | 48 | 33 | 27 | 21 | 17 | 13 | 5 | 4 | 20 |
| Australia | 24 | 33 | 54 | 91 | 89 | 89 | 21 | 28 | 28 | 31 | 25 | 20 | 7 | 6 | 30 |
| Hungary | 38 | 20 | 7 | 19 | 5 | 31 | 49 | 30 | 29 | 24 | 19 | 14 | 7 | 5 | 19 |
| Turkey | 42 | 52 | 42 | 23 | 49 | 29 | 15 | 27 | 30 | 30 | 21 | 19 | 6 | 6 | 34 |
| Norway | 15 | 22 | 52 | 100 | 36 | 112 | 42 | 43 | 31 | 34 | 27 | 22 | 9 | 6 | 32 |
| Slovenia | 23 | 14 | 12 | 20 | 18 | 19 | 57 | 48 | 32 | 27 | 20 | 16 | 5 | 5 | 23 |
| Brazil | 57 | 72 | 34 | 64 | 59 | 72 | 11 | 23 | 33 | 36 | 29 | 24 | 7 | 6 | 35 |
| Portugal | 32 | 34 | 55 | 69 | 51 | 22 | 44 | 41 | 34 | 33 | 26 | 22 | 8 | 6 | 44 |
| Argentina | 31 | 62 | 45 | 43 | 45 | 84 | 17 | 42 | 35 | 32 | 24 | 21 | 8 | 6 | 38 |
| Russia | 60 | 57 | 53 | 39 | 79 | 100 | 18 | 20 | 36 | 38 | 30 | 26 | 7 | 6 | 39 |
| Saudi Arabia | 39 | 46 | 23 | 80 | 61 | 116 | 30 | 35 | 37 | 35 | 28 | 23 | 8 | 7 | 40 |
| Indonesia | 77 | 85 | 30 | 11 | 68 | 80 | 16 | 29 | 38 | 38 | 36 | 21 | 8 | 6 | 36 |
| Kuwait | 26 | 25 | 75 | 89 | 103 | 94 | 55 | 46 | 39 | 45 | 35 | 31 | 10 | 7 | 51 |
| Belarus | 50 | 41 | 73 | 3 | 54 | 26 | 51 | 47 | 40 | 36 | 28 | 25 | 8 | 6 | 46 |
| South Africa | 58 | 58 | 61 | 54 | 43 | 70 | 33 | 37 | 41 | 39 | 31 | 26 | 8 | 6 | 43 |
| Luxembourg | 16 | 3 | 117 | 113 | 56 | 35 | 77 | 62 | 42 | 48 | 41 | 28 | 10 | 6 | 42 |
| India | 103 | 104 | 32 | 51 | 70 | 38 | 9 | 15 | 43 | 42 | 35 | 26 | 7 | 6 | 37 |

| Country | | | | | | | | | | | | | | | |
|---|---|---|---|---|---|---|---|---|---|---|---|---|---|---|---|
| Philippines | 79 | 79 | 15 | 18 | 3 | 8 | 34 | 38 | 44 | 32 | 23 | 22 | 8 | 5 | 31 |
| Chile | 46 | 48 | 71 | 50 | 107 | 88 | 43 | 44 | 45 | 44 | 34 | 30 | 10 | 7 | 59 |
| Romania | 75 | 45 | 37 | 66 | 32 | 21 | 54 | 39 | 46 | 37 | 30 | 25 | 8 | 6 | 45 |
| Lithuania | 47 | 30 | 74 | 32 | 57 | 36 | 69 | 51 | 47 | 43 | 33 | 29 | 10 | 7 | 50 |
| New Zealand | 29 | 37 | 87 | 70 | 86 | 91 | 52 | 57 | 48 | 50 | 40 | 33 | 12 | 6 | 52 |
| Greece | 36 | 50 | 78 | 101 | 58 | 58 | 46 | 54 | 49 | 46 | 38 | 29 | 11 | 8 | 48 |
| Croatia | 44 | 42 | 40 | 44 | 38 | 20 | 61 | 64 | 50 | 40 | 32 | 27 | 8 | 7 | 49 |
| Venezuela | 51 | 66 | 36 | 41 | 117 | 108 | 36 | 49 | 50 | 48 | 37 | 32 | 10 | 7 | 62 |
| Estonia | 45 | 19 | 46 | 49 | 50 | 34 | 82 | 63 | 51 | 42 | 35 | 26 | 10 | 7 | 41 |
| Ukraine | 88 | 59 | 63 | 22 | 47 | 37 | 50 | 40 | 52 | 41 | 32 | 28 | 10 | 6 | 53 |
| Vietnam | 96 | 78 | 66 | 12 | 72 | 68 | 45 | 36 | 53 | 53 | 43 | 34 | 11 | 7 | 66 |
| Iran | 72 | 86 | 24 | 42 | 81 | 104 | 35 | 45 | 54 | 52 | 42 | 33 | 11 | 7 | 58 |
| Costa Rica | 41 | 51 | 80 | 24 | 23 | 59 | 60 | 70 | 55 | 47 | 38 | 30 | 8 | 7 | 47 |
| Qatar | 28 | 17 | 77 | 130 | 71 | 124 | 78 | 65 | 56 | 60 | 45 | 41 | 11 | 6 | 55 |
| Tunisia | 62 | 53 | 99 | 38 | 46 | 43 | 58 | 58 | 57 | 49 | 39 | 33 | 10 | 6 | 54 |
| Bulgaria | 69 | 47 | 47 | 48 | 62 | 65 | 70 | 56 | 58 | 51 | 41 | 33 | 11 | 7 | 57 |
| Trinidad and Tobago | 52 | 29 | 26 | 103 | 92 | 57 | 84 | 68 | 58 | 55 | 46 | 36 | 12 | 9 | 56 |
| Malta | 37 | 18 | 16 | 86 | 27 | 9 | 92 | 78 | 59 | 33 | 26 | 22 | 8 | 6 | 33 |
| Egypt | 71 | 100 | 56 | 34 | 74 | 76 | 32 | 53 | 60 | 54 | 44 | 35 | 11 | 7 | 68 |
| Peru | 66 | 75 | 84 | 59 | 124 | 85 | 47 | 50 | 61 | 61 | 48 | 40 | 12 | 7 | 75 |
| Colombia | 67 | 93 | 65 | 67 | 60 | 107 | 41 | 59 | 62 | 61 | 47 | 41 | 12 | 7 | 76 |
| Iceland | 12 | 35 | 86 | 84 | 44 | 113 | 83 | 90 | 63 | 69 | 53 | 45 | 12 | 7 | 77 |
| Morocco | 84 | 81 | 57 | 68 | 55 | 50 | 53 | 57 | 64 | 59 | 45 | 40 | 11 | 6 | 73 |
| Hong Kong (China) | 64 | 56 | 39 | 132 | 35 | 83 | 66 | 66 | 65 | 57 | 46 | 37 | 11 | 7 | 79 |
| Latvia | 63 | 38 | 64 | 96 | 63 | 45 | 85 | 69 | 66 | 59 | 46 | 39 | 11 | 9 | 65 |
| Oman | 48 | 49 | 79 | 104 | 48 | 122 | 73 | 72 | 67 | 67 | 50 | 44 | 13 | 10 | 80 |
| Kazakhstan | 74 | 65 | 106 | 65 | 52 | 118 | 56 | 61 | 68 | 63 | 47 | 43 | 13 | 6 | 78 |
| El Salvador | 59 | 77 | 70 | 16 | 97 | 25 | 65 | 76 | 69 | 62 | 50 | 40 | 11 | 7 | 67 |
| Jordan | 68 | 67 | 48 | 40 | 41 | 48 | 73 | 74 | 70 | 58 | 46 | 38 | 11 | 7 | 60 |
| Uruguay | 35 | 74 | 88 | 57 | 85 | 98 | 61 | 81 | 71 | 68 | 51 | 46 | 13 | 9 | 81 |
| Pakistan | 104 | 110 | 51 | 35 | 111 | 44 | 40 | 52 | 72 | 56 | 45 | 37 | 11 | 7 | 69 |
| Lebanon | 56 | 68 | 69 | 99 | 42 | 62 | 72 | 79 | 73 | 65 | 51 | 42 | 11 | 7 | 61 |
| Serbia | 99 | 64 | 68 | 46 | 64 | 49 | 81 | 67 | 74 | 66 | 48 | 44 | 11 | 7 | 71 |
| Guatemala | 86 | 84 | 81 | 74 | 88 | 67 | 68 | 71 | 75 | 73 | 54 | 50 | 13 | 7 | 88 |
| Bangladesh | 107 | 111 | 67 | 37 | 127 | 12 | 46 | 60 | 76 | 66 | 53 | 41 | 12 | 6 | 70 |
| Mauritius | 55 | 55 | 124 | 47 | 131 | 5 | 86 | 88 | 77 | 71 | 58 | 44 | 14 | 11 | 63 |
| Sri Lanka | 94 | 89 | 92 | 61 | 113 | 66 | 64 | 71 | 78 | 77 | 59 | 52 | 14 | 10 | 91 |
| Syria | 90 | 98 | 59 | 58 | 84 | 92 | 62 | 73 | 79 | 74 | 56 | 50 | 14 | 7 | 85 |
| Algeria | 100 | 83 | 94 | 116 | 133 | 114 | 59 | 55 | 80 | 79 | 61 | 53 | 14 | 11 | 89 |
| Bosnia and Herzegovina | 89 | 60 | 43 | 90 | 83 | 61 | 88 | 78 | 81 | 70 | 52 | 48 | 13 | 9 | 83 |
| FYR Macedonia | 70 | 63 | 83 | 33 | 91 | 74 | 88 | 85 | 82 | 72 | 55 | 48 | 13 | 10 | 84 |
| Swaziland | 61 | 61 | 132 | 4 | 69 | 10 | 91 | 91 | 83 | 64 | 49 | 43 | 11 | 10 | 64 |
| Botswana | 95 | 43 | 58 | 127 | 125 | 7 | 94 | 75 | 84 | 76 | 63 | 49 | 14 | 11 | 82 |
| Ecuador | 83 | 92 | 102 | 63 | 82 | 117 | 67 | 77 | 85 | 78 | 60 | 53 | 14 | 11 | 93 |
| Cyprus | 49 | 73 | 91 | 114 | 21 | 55 | 88 | 94 | 86 | 73 | 57 | 47 | 14 | 10 | 72 |
| Côte d'Ivoire | 106 | 102 | 82 | 36 | 65 | 99 | 75 | 77 | 87 | 80 | 62 | 53 | 12 | 7 | 86 |
| Cambodia | 105 | 97 | 131 | 27 | 118 | 73 | 80 | 77 | 88 | 83 | 66 | 54 | 15 | 13 | 90 |
| Honduras | 80 | 106 | 104 | 26 | 73 | 95 | 74 | 89 | 89 | 79 | 61 | 53 | 14 | 12 | 92 |
| Bolivia | 97 | 91 | 116 | 62 | 130 | 96 | 79 | 80 | 90 | 84 | 66 | 56 | 15 | 13 | 94 |
| Jamaica | 81 | 82 | 72 | 108 | 123 | 11 | 89 | 90 | 91 | 81 | 64 | 52 | 14 | 13 | 95 |
| Albania | 87 | 88 | 85 | 85 | 94 | 54 | 89 | 90 | 92 | 82 | 62 | 55 | 14 | 10 | 103 |
| Nigeria | 125 | 108 | 38 | 125 | 120 | 119 | 63 | 50 | 93 | 88 | 70 | 59 | 15 | 11 | 87 |
| Georgia | 102 | 99 | 60 | 88 | 39 | 53 | 91 | 92 | 94 | 84 | 65 | 57 | 15 | 13 | 96 |
| Cameroon | 101 | 113 | 96 | 28 | 108 | 105 | 71 | 89 | 95 | 86 | 69 | 58 | 15 | 13 | 102 |
| Armenia | 91 | 101 | 110 | 56 | 78 | 69 | 91 | 93 | 96 | 85 | 67 | 58 | 15 | 13 | 104 |
| Paraguay | 93 | 105 | 89 | 76 | 104 | 119 | 83 | 92 | 97 | 87 | 70 | 58 | 15 | 13 | 98 |
| Congo | 110 | 76 | 125 | 120 | 1 | 101 | 95 | 83 | 98 | 90 | 70 | 60 | 16 | 14 | 100 |

| Country | | | | | | | | | | | | | | | |
|---|---|---|---|---|---|---|---|---|---|---|---|---|---|---|---|
| Kenya | 115 | 114 | 119 | 93 | 77 | 86 | 76 | 82 | 99 | 91 | 71 | 62 | 15 | 13 | 97 |
| Senegal | 112 | 107 | 76 | 87 | 101 | 64 | 89 | 87 | 99 | 89 | 68 | 61 | 15 | 13 | 108 |
| Barbados | 78 | 69 | 29 | 128 | 53 | 16 | 98 | 97 | 100 | 75 | 55 | 51 | 11 | 10 | 74 |
| Gabon | 92 | 70 | 113 | 126 | 110 | 121 | 95 | 90 | 100 | 93 | 72 | 63 | 16 | 13 | 109 |
| Fiji | 82 | 87 | 112 | 81 | 112 | 82 | 96 | 96 | 101 | 92 | 72 | 62 | 15 | 15 | 106 |
| Tanzania | 116 | 117 | 129 | 92 | 102 | 87 | 75 | 86 | 102 | 96 | 73 | 65 | 16 | 13 | 112 |
| Azerbaijan | 109 | 95 | 108 | 129 | 93 | 128 | 91 | 84 | 103 | 94 | 72 | 64 | 16 | 15 | 115 |
| Suriname | 76 | 71 | 93 | 94 | 114 | 123 | 97 | 96 | 104 | 94 | 74 | 61 | 16 | 13 | 99 |
| Mongolia | 111 | 80 | 114 | 105 | 132 | 75 | 97 | 89 | 105 | 97 | 73 | 66 | 16 | 15 | 111 |
| Panama | 73 | 112 | 109 | 117 | 96 | 106 | 83 | 97 | 106 | 95 | 74 | 62 | 16 | 14 | 101 |
| Zambia | 117 | 109 | 62 | 93 | 100 | 120 | 91 | 88 | 107 | 95 | 72 | 65 | 16 | 13 | 113 |
| Macao (China) | 53 | 94 | 122 | 131 | 121 | 93 | 93 | 98 | 108 | 105 | 80 | 70 | 17 | 16 | 118 |
| Belize | 65 | 90 | 74 | 77 | 134 | 110 | 97 | 98 | 109 | 98 | 76 | 64 | 17 | 15 | 107 |
| Moldova | 113 | 103 | 111 | 98 | 105 | 79 | 96 | 94 | 110 | 99 | 75 | 66 | 17 | 15 | 110 |
| Tajikistan | 108 | 124 | 126 | 5 | 12 | 126 | 91 | 97 | 111 | 104 | 82 | 68 | 17 | 16 | 122 |
| Madagascar | 122 | 118 | 123 | 73 | 128 | 63 | 91 | 93 | 112 | 101 | 77 | 67 | 17 | 15 | 116 |
| Kyrgyzstan | 118 | 115 | 118 | 82 | 90 | 115 | 96 | 96 | 113 | 103 | 78 | 70 | 18 | 16 | 121 |
| Ghana | 123 | 119 | 130 | 106 | 76 | 127 | 90 | 93 | 114 | 106 | 78 | 71 | 17 | 16 | 129 |
| Nepal | 127 | 121 | 127 | 110 | 87 | 51 | 91 | 93 | 114 | 103 | 80 | 69 | 17 | 16 | 117 |
| Uganda | 124 | 125 | 95 | 109 | 95 | 102 | 87 | 95 | 115 | 102 | 79 | 68 | 17 | 16 | 114 |
| Yemen | 121 | 120 | 121 | 118 | 122 | 130 | 88 | 94 | 116 | 107 | 80 | 72 | 17 | 16 | 128 |
| Mozambique | 114 | 127 | 97 | 79 | 115 | 131 | 85 | 97 | 117 | 107 | 81 | 71 | 17 | 16 | 124 |
| Saint Lucia | 85 | 96 | 103 | 124 | 67 | 78 | 98 | 99 | 117 | 100 | 78 | 65 | 17 | 15 | 105 |
| Cape Verde | 98 | 116 | 44 | 102 | 135 | 81 | 98 | 99 | 118 | 107 | 81 | 71 | 19 | 16 | 127 |
| Malawi | 128 | 123 | 90 | 97 | 99 | 111 | 95 | 96 | 119 | 108 | 82 | 73 | 19 | 16 | 119 |
| Haiti | 120 | 130 | 115 | 95 | 129 | 42 | 94 | 98 | 120 | 111 | 85 | 75 | 19 | 18 | 125 |
| Sudan | 119 | 129 | 101 | 112 | 126 | 132 | 80 | 96 | 120 | 110 | 84 | 74 | 19 | 17 | 120 |
| Niger | 132 | 122 | 49 | 121 | 106 | 71 | 97 | 96 | 121 | 109 | 83 | 73 | 17 | 16 | 123 |
| Rwanda | 126 | 126 | 107 | 115 | 119 | 90 | 96 | 98 | 122 | 113 | 86 | 77 | 20 | 19 | 130 |
| Ethiopia | 131 | 133 | 98 | 123 | 66 | 129 | 88 | 97 | 123 | 112 | 86 | 76 | 17 | 16 | 126 |
| CAR | 130 | 131 | 100 | 111 | 116 | 109 | 98 | 99 | 124 | 115 | 87 | 79 | 21 | 21 | 132 |
| Burundi | 133 | 134 | 128 | 107 | 80 | 125 | 98 | 99 | 125 | 116 | 88 | 80 | 22 | 22 | 135 |
| Eritrea | 134 | 135 | 105 | 119 | 98 | 103 | 98 | 99 | 126 | 117 | 89 | 80 | 23 | 23 | 134 |
| Gambia | 129 | 128 | 120 | 122 | 109 | 97 | 99 | 99 | 126 | 114 | 87 | 78 | 21 | 20 | 131 |
| Iraq | 135 | 132 | 50 | 133 | 75 | 133 | 97 | 98 | 126 | 115 | 87 | 79 | 21 | 20 | 133 |

**Andrey Subochev**

DeCAn Lab and Department of Mathematics for Economics, National Research University Higher School of Economics, Moscow, asubochev@hse.ru

**Igor Zakhlebin**

National Research University Higher School of Economics, Moscow, zahl.igor@gmail.com

**Название на русском:**

Альтернативные версии глобального рейтинга конкурентоспособности промышленного производства, построенные методами теории коллективного выбора

**Имена авторов на русском:**

А.Н. Субочев

И.В. Захлебин

Национальный исследовательский университет "Высшая школа экономики"

**Аннотация на русском:** Индекс конкурентоспособности промышленного производства, разработанный экспертами ЮНИДО, предназначен служить мерой национальной конкурентоспособности. Индекс является агрегатом восьми наблюдаемых переменных, с разных сторон характеризующие результативность промышленного производства. Вместо того, чтобы использовать кардинальную агрегирующую функцию, как это делают авторы индекса, предлагается применить ординальные методы ранжирования, заимствованные из теории коллективного выбора, основанные на правиле большинства, такие как правило Коупланда, марковский метод и многоступенчатой процедура отбора наилучших альтернатив, определяемых с помощью решений задачи оптимального коллективного выбора (турнирных решений), таких как непокрытое множество и минимальное внешнеустойчивое множество. Тот же самый метод парных сравнений с помощью правила большинства используется для анализа ранговых корреляций. Показано, что некоторые из новых версий глобального рейтинга представляют данный набор критериев лучше, чем исходная версия, основанная на индексе.